\documentclass{aamas2010_extendedAbstract}

\usepackage{algorithm}
\usepackage{algorithmic}
\usepackage{multirow} 
\usepackage{accents}

\newtheorem{theorem}{Theorem}
 
\newdef{definition}{Definition}

\begin{document}



\AuthorsForCitationInfo{Arthur Carvalho, Kate Larson}

\TitleForCitationInfo{Sharing a Reward Based on Peer Evaluations}

\title{Sharing a Reward Based on Peer Evaluations}

\numberofauthors{2}

\author{
\alignauthor
Arthur Carvalho\\
       \affaddr{Cheriton School of Computer Science}\\
       \affaddr{University of Waterloo}\\
       \affaddr{Waterloo, ON, Canada}\\
       \email{a3carval@cs.uwaterloo.ca}
\alignauthor
Kate Larson\\
       \affaddr{Cheriton School of Computer Science}\\
       \affaddr{University of Waterloo}\\
       \affaddr{Waterloo, ON, Canada}\\
       \email{klarson@cs.uwaterloo.ca}
}

\maketitle

\begin{abstract}
We study a problem where a group of agents has to decide how some fixed value should be shared among them. We are interested in settings where the share that each agent receives is based on how that agent is evaluated by other members of the group, where highly regarded agents receive a greater share compared to agents that are not well regarded. We introduce two mechanisms for determining agents' shares: the \emph{peer-evaluation mechanism}, where each agent gives a direct evaluation for every other member of the group, and the \emph{peer-prediction mechanism}, where each agent is asked to report how they believe group members will evaluate a particular agent. The sharing is based on the provided information. While both mechanisms are individually rational, the first mechanism is strategy-proof and budget-balanced, but it can be collusion-prone. Further, the second mechanism is collusion-resistant and incentive-compatible.
\end{abstract}




\section{Model and Background}

Let $N = \{1,\dots,n\}$ be a set of \emph{agents} that must share a \emph{reward} $V \in \Re^+$. We are interested in settings where the share of $V$ that an agent receives depends on evaluations that its peers make concerning the agent's contribution to the group. Hence, each agent $i\in N$ is asked to provide \emph{evaluations} for all peers. Such evaluations can be either \emph{direct evaluations} or \emph{predictions} of absolute frequencies of received evaluations. For avoiding a biased self-evaluation, an agent is not requested to provide evaluations for itself.

Given a positive integer parameter $0 < M \leq V$, the direct evaluations made by an agent $i \in N$ are formally represented by the vector $\textbf{t}_i = (t_{i}^1,\dots,t_{i}^{i-1},t_{i}^{i+1},\dots, t_{i}^n)$, where $t_i^j\in \{0,\dots,M\}$ represents agent $i$'s evaluation given to agent $j$, and $\sum_{j\neq i} t_i^j= M$. Hence, the parameter $M$ represents the top possible evaluation that an agent can receive and an explicit constraint that bounds the sum of direct  evaluations.

The predictions made by agent $i$ are formally represented by the vector $\textbf{r}_i= (r_{i}^1,\dots,r_{i}^{i-1},r_{i}^{i+1},\dots, r_{i}^n)$, where $\textbf{r}_{i}^j = (r_{i}^{j^0},\dots, r_{i}^{j^M})$ represents the agent $i$'s prediction for the absolute frequency of evaluations given to agent $j$, \textit{i.e.} $r_{i}^{j^k} \in \{1,\dots, n-1 \},$ for $0 \leq  k \leq M$, and  $\sum_{k=0}^M r_{i}^{j^k} = n-1$.

The evaluations are submitted to a central entity called \emph{mechanism}, which is responsible for sharing the reward. This entity relies only on reported evaluations when determining agents' shares. We assume that evaluations are independent across agents, that evaluations provided by an agent for its peers are independent among themselves, and that agents act to maximize their expected shares. This implies that agents may deliberately lie when providing evaluations for others. Therefore, we distinguish between the true evaluations made by agent $i$, $\textbf{t}_i$ for direct evaluations and $\textbf{r}_i$ for predictions, and the evaluations that it reports to the mechanism, $\textbf{x}_i=(x_{i}^1,$ $\dots,$ $x_{i}^{i-1}, x_{i}^{i+1}, \ldots,   x_{i}^n)$. We overload the notation using $\textbf{x}_i$ to denote both direct evaluations and predictions, but we make clear its meaning when necessary.

We  call  $\textbf{x}_i$ the \emph{strategy} of agent $i$ and $\textbf{X} = (\textbf{x}_1, \ldots, \textbf{x}_n)$ a \emph{strategy profile}. We define $\textbf{X}_{-i}=(\textbf{x}_1,\ldots, \textbf{x}_{i-1}, \textbf{x}_{i+1}, $ $\ldots, \textbf{x}_n)$. Thus, we can represent a strategy profile as $\textbf{X} = (\textbf{x}_i,\textbf{X}_{-i})$. If the reported evaluation of agent $i$ is equal to its true evaluation, \textit{i.e.} $\textbf{x}_i=\textbf{t}_i$ for direct evaluations or $\textbf{x}_i=\textbf{r}_i$ for predictions, then we say that agent $i$'s strategy is \emph{truthful}, and represent it by $\accentset{*}{\textbf{x}}_i$. We say that $\textbf{X}$ is \emph{collectively truthful} if all reported strategies are truthful. We denote the share of $V$ given to agent $i$ when all the reported evaluations are $\textbf{X}$ by $\Gamma_i(\textbf{X})$. The most important property we wish our mechanisms to have is that the share assigned to each agent should reflect the reported evaluations for that agent. In addition to this requirement, we would like our mechanisms to be \emph{budget-balanced}, \emph{individually rational}, \emph{incentive-compatible} (or \emph{strategy-proof}), and \emph{collusion-resistant} \cite{Myerson}. We consider that a \emph{collusion} between agents $i$ and $j$ occurs when agent $i$ changes its truthful evaluation for agent $j$, resulting in the report $\hat{\textbf{x}}_i \neq \accentset{*}{\textbf{x}}_i$, and, for doing this, it receives a side-payment,  $p$, so that: 

\begin{enumerate}
    \item $\mathbb{E} \left[ \Gamma_i(\hat{\textbf{x}}_i, \textbf{X}_{-i}) \right] + p > \mathbb{E} \left[ \Gamma_i(\accentset{*}{\textbf{x}}_i , \textbf{X}_{-i})\right]$;
\item $\mathbb{E} \left[ \Gamma_j(\hat{\textbf{x}}_i,\textbf{X}_{-i})\right]-p> \mathbb{E} \left[\Gamma_j(\accentset{*}{\textbf{x}}_i,\textbf{X}_{-i})\right]$.
\end{enumerate}

Collusions with more than two agents can be decomposed into a union of collusions between two agents (a liar and a beneficiary). We say that a mechanism is \emph{collusion-resistant} when, for all agents $i,j \in N$ and strategies $\hat{\textbf{x}}_i \neq \accentset{*}{\textbf{x}}_i$, where $\hat{x}^{j}_i > \accentset{*}{x}^j_i$ for direct evaluations and  $\sum_{k = 0}^M k(\hat{x}^{j^k}_i - \accentset{*}{x}^{j^k}_i) > 0$ for predictions, we have $\mathbb{E} \left[ \Gamma_i(\hat{\textbf{x}}_i,\textbf{X}_{-i}) + \Gamma_j(\hat{\textbf{x}}_i,\textbf{X}_{-i}) \right] \leq
\mathbb{E} \left[ \Gamma_i(\accentset{*}{\textbf{x}}_i,\textbf{X}_{-i}) + \Gamma_j(\accentset{*}{\textbf{x}}_i,\textbf{X}_{-i}) \right]$. To provide incentives for truth-telling,  we use the following strictly proper scoring rule \cite{proper-scoring-rules}:

\begin{equation}
R(\textbf{p},e) = 1 + 2p_e - \sum_{j=1}^z p_{j}^2 \in [0,\,2]
\end{equation}
where $\textbf{p}$ is a probability distribution and $e$ is the observed event among $z$ possible outcomes.

\section{The Peer-Evaluation Mechanism}

The peer-evaluation mechanism announces the parameter $M$ and requests agents to submit direct evaluations. The sharing scheme is presented in Algorithm 1. The share received by each agent $i \in N$ is computed by aggregating its received evaluations into a variable $grade_i$, and multiplying it by a normalizing factor $V/(n\times M)$. Due to the constraint imposed on direct evaluations, \textit{i.e.} $\sum_{j \neq i} x_i^j = M$, it is clear that after this operation $\sum_{i = 1}^n \Gamma_i = V$. Consequently, the mechanism is budget-balanced. Because the evaluations are greater than or equal to zero, an agent cannot receive a negative share. Then, the mechanism is individually rational. The following theorem states our main result concerning the properties of the peer-evaluation mechanism.

\begin{theorem}
The peer-evaluation mechanism is strategy-proof.
\end{theorem}

\begin{algorithm}[b]
\caption{The Peer-Evaluation Mechanism}
\begin{algorithmic}[1] 
\FOR{$i = 1$ to $n$} 
\STATE $grade_i = \sum_{j \neq i} x_j^i$
\STATE $\Gamma_i = grade_i\times \frac{V}{n  M}$
\ENDFOR
\end{algorithmic}
\end{algorithm}

The main drawback of the peer-evaluation mechanism is that agents do not have direct incentives for lying, but they also do not have incentives for telling the truth. This characteristic makes the mechanism extremely susceptible to collusions.

\section{The Peer-Prediction Mechanism}

The peer-prediction mechanism announces the parameter $M$ and requests agents to submit predictions. We can see this game as if each agent $i$ was answering the following question about each other agent $j$: \textquotedblleft if agents were to evaluate agent $j$, what would be the absolute frequency of the evaluations received by it?\textquotedblright. The sharing scheme is presented in Algorithm 2. The main idea of the peer-prediction mechanism is to compute agents' shares using \emph{grades}, which are aggregations of the expected evaluations calculated from predictions, and using \textit{scoring rules} \cite{proper-scoring-rules} to generate \emph{scores} and enforce truth-telling. For using scoring rules, it is necessary to have a \textquotedblleft reality\textquotedblright\space to score an assessment. Our solution considers grades as observed events of an uncertain quantity, with possible outcomes inside the set $\{0, \dots,M\}$, and scores the reported predictions as if they were assessments

\begin{algorithm}[t]
\caption{The Peer-Prediction Mechanism}
\begin{algorithmic}[1]                    

\FOR{$i = 1$ to $n$} 

\STATE $g_i = \sum_{j \neq i}\sum_{k=0}^M \frac{x_j^{i^k}}{n-1}\times k$

\ENDFOR

\FOR{$i = 1$ to $n$} 
\STATE $score_i = \frac{\sum_{j \neq i} R\left(\frac{x_i^j}{n-1}, \, nint\left(\frac{g_j - \sum_{k=0}^M \frac{x_i^{j^k}}{n-1}\times k}{n-2}\right)\right)}{n-1}$

\STATE $grade_i = \frac{g_i}{n-1}$

\STATE $\Gamma_i  = (grade_i + \alpha\times score_i)\times \frac{V}{(M + 2\alpha)n}$
\ENDFOR

\end{algorithmic}
\end{algorithm}

The sharing process has essentially four steps. The first one transforms all the predictions about the evaluations for an agent $i \in N$ to a positive real number, $g_i$, by creating a probability distribution from each prediction $x_*^i$, and summing the expected value of each distribution.

In the second step, the score of each agent $i \in N$ is calculated as follows: First, for each agent $j \neq i$, a probability distribution is created from the prediction $x_i^j$. Second, a temporary grade for agent $j$ is calculated as the arithmetic mean of the expected evaluations received by it, without taking into consideration the expected value from agent $i$'s prediction. The function $nint$ (nearest integer function) rounds this temporary grade to an integer number inside the set $\{0, \dots, M\}$. Finally, the mechanism applies the strictly proper scoring rule represented by Equation 1 on the probability distribution (assessment) and the temporary grade (observed event). In the end, the score of agent $i$ is the arithmetic mean of results provided by the scoring rule for each prediction submitted by agent $i$.

In the third step, agents' grades are computed as the arithmetic mean of the expected evaluations received by them. Finally, they have their shares computed in the last step. Agents' scores are multiplied by a constant $\alpha > 0$ and added to their grades. The result is then multiplied by a weight $V/(M + 2\alpha)n$ to form agents' shares. The constant $\alpha$ fine-tunes the weight given to scores. Because the highest grade that an agent can receive is equal to $M$, and the highest score is equal to $2\alpha$, using the weight $V/(M + 2\alpha )n$ guarantees that  the mechanism will not make a loss in the case of every agent receives the highest grade and score. An obvious consequence of such approach is that when at least one agent does not receive the highest grade or score, then the mechanism will make a profit. This implies that it is not always budget-balanced. Given that scores and grades are always greater than or equal to zero, the mechanism is individually rational. The following theorem states our main result related to the  peer-prediction mechanism.

\begin{theorem}
The peer-prediction mechanism is incentive-compatible.
\end{theorem}

Related to collusions, we have the following result:

\begin{theorem}
If $\alpha > M(n-1)/2$, then the peer-prediction mechanism is collusion-resistant.
\end{theorem}

\bibliographystyle{abbrv}

\end{document}